\documentclass[preprint,12pt]{elsarticle}

\usepackage{amssymb}
\journal{Nuclear Instruments and Methods in Physics Research A}

\begin{document}

\begin{frontmatter}


\title{Positron production using a 9 MeV electron linac for the GBAR experiment}

\author[label1]{M.~Charlton}
\author[label2]{J.~J.~Choi}
\author[label3]{M.~Chung}
\author[label4]{P.~Clad\'{e}}
\author[label5]{P.~Comini}
\author[label4]{P-P.~Cr\'{e}pin}
\author[label6]{P.~Crivelli}
\author[label7]{O.~Dalkarov}
\author[label5]{P.~Debu}
\author[label1]{L.~Dodd}
\author[label4,label4a]{A.~Douillet}
\author[label4]{S.~Guellati-Kh\'{e}lifa}
\author[label8]{P-A.~Hervieux}
\author[label4,label4a]{L.~Hilico}
\author[label9]{A.~Husson\fnref{AH}}
\author[label4]{P.~Indelicato}
\author[label6]{G.~Janka}
\author[label10]{S.~Jonsell}
\author[label4,label4a]{J-P.~Karr}
\author[label2]{B.~H.~Kim}
\author[label11]{E-S.~Kim}
\author[label2]{S.~K.~Kim}
\author[label12]{Y.~Ko}
\author[label13]{T.~Kosinski}
\author[label14]{N.~Kuroda}
\author[label5]{B.~Latacz\fnref{BL}}
\author[label2]{H.~Lee}
\author[label12]{J.~Lee}
\author[label5]{A.~M.~M.~Leite\fnref{AL}}
\author[label8]{K.~L\'{e}v\^{e}que}
\author[label11]{E.~Lim}
\author[label5]{L.~Liszkay\corref{LL}}
\ead{laszlo.liszkay@cea.fr}
\author[label5]{P.~Lotrus}
\author[label4]{T.~Louvradoux}
\author[label9]{D.~Lunney}
\author[label8]{G.~Manfredi}
\author[label5]{B.~Mansouli\'{e}}
\author[label13]{M.~Matusiak}
\author[label15]{G.~Mornacchi}
\author[label16]{V. V.~Nesvizhevsky}
\author[label4]{F.~Nez}
\author[label5]{S.~Niang}
\author[label14]{R.~Nishi}
\author[label15]{S.~Nourbaksh}
\author[label2]{K.~H.~Park}
\author[label4]{N.~Paul}
\author[label5]{P.~P\'{e}rez}
\author[label5]{S.~Procureur}
\author[label6]{B.~Radics}
\author[label6]{C.~Regenfus}
\author[label5]{J-M.~Rey\fnref{JMR}}
\author[label5]{J-M.~Reymond}
\author[label4]{S.~Reynaud}
\author[label5]{J-Y.~Rouss\'{e}}
\author[label4]{O.~Rousselle}
\author[label6]{A.~Rubbia}
\author[label13]{J.~Rzadkiewicz}
\author[label5]{Y.~Sacquin}
\author[label17]{F.~Schmidt-Kaler}
\author[label13]{M.~Staszczak}
\author[label5]{B.~Tuchming}
\author[label5]{B.~Vallage}
\author[label7]{A.~Voronin}
\author[label15]{A.~Welker}
\author[label1]{D. P.~van der Werf}
\author[label17]{S.~Wolf}
\author[label2]{D.~Won}
\author[label13]{S.~Wronka}
\author[label18]{Y.~Yamazaki}
\author[label3]{K-H.~Yoo}

\address[label1]{Department of Physics, College of Science, Swansea University, Swansea SA2 8PP, United Kingdom}
\address[label2]{Department of Physics and Astronomy, Seoul National University, 599 Gwanak-Ro, Gwanak-gu, Seoul 08826, Korea}
\address[label3]{Department of Physics, Ulsan National Institute of Science and Technology (UNIST), 50, UNIST-gil, Ulsan 44919, Republic of Korea}
\address[label4]{Laboratoire Kastler Brossel, Sorbonne Universit\'e, CNRS, ENS-PSL Research University, Coll\`ege de France, Case\ 74;\ 4, place Jussieu, F-75005 Paris, France}
\address[label4a]{Universit\'{e} d'\'Evry-Val d'Essonne, Universit\'e Paris-Saclay, Boulevard François Mitterrand, F-91000 \'Evry, France}
\address[label5]{IRFU, CEA, Universit\'{e} Paris-Saclay, F-91191 Gif-sur-Yvette Cedex, France}
\address[label6]{Institute for Particle Physics and Astrophysics, ETH Zürich, CH-8093 Zürich, Switzerland}
\address[label7]{P. N. Lebedev Physical Institute, 53 Leninsky Prospect, 117991 Moscow, Russia}
\address[label8]{Universit\'{e} de Strasbourg, CNRS, Institut de Physique et Chimie des Mat\'{e}riaux de Strasbourg, UMR 7504, F-67000 Strasbourg, France}
\address[label9]{ Universit\'{e} Paris Sud - Paris 11, CSNSM IN2P3 CNRS, Orsay, France}
\address[label10]{Department of Physics, Stockholm University, SE-10691 Stockholm, Sweden}
\address[label11]{Department of Accelerator Science, Korea University Sejong Campus, Sejong-ro 2511, 0019 Sejong, Republic of Korea}
\address[label12]{Center for Underground Physics, Institute for  Basic Science, 70 Yuseong-daero 1689-gil,  Yuseong-gu, Daejeon 34047, Korea}
\address[label13]{National Centre for Nuclear Research (NCBJ), ul. Andrzeja Sołtana 7, 05-400 Otwock, Swierk, Poland}
\address[label14]{Institute of Physics, University of Tokyo, 3-8-1 Komaba, Meguro, Tokyo 153-8902, Japan}
\address[label15]{CERN, 1211 Geneva 23, Switzerland}
\address[label16]{Institut Max von Laue - Paul Langevin (ILL), 71 avenue des Martyrs, F-38042 Grenoble, France}
\address[label17]{QUANTUM, Institut für Physik, Johannes Gutenberg Universit\"{a}t, D-55128 Mainz, Germany}
\address[label18]{Ulmer Fundamental Symmetries Laboratory, RIKEN, 2-1 Hirosawa, Wako, 351-0198, Saitama, Japan}

\cortext[LL]{Corresponding author.}
\fntext[JMR]{present address: CEA Saclay, POSITH\^OT, 91191 Gif sur Yvette, France}
\fntext[BL]{present address: RIKEN, Ulmer Fundamental Symmetries Laboratory, Wako, Saitama 351-0198, Japan}
\fntext[AL]{present address: Institut Curie, PSL Research University, Radiation Oncology
Department, Proton Therapy Centre, Centre Universitaire, 91898, Orsay,
France}
\fntext[AH]{present address: CENBG, 19 Chemin du Solarium, CS 10120, F-33175 Grandignan Cedex, France}
\begin{abstract}
For the GBAR (Gravitational Behaviour of Antihydrogen at Rest) experiment at CERN's Antiproton Decelerator (AD) facility we have constructed a source of slow positrons, which uses a low-energy electron linear accelerator (linac).  The driver linac produces electrons of 9~MeV kinetic energy that create positrons from bremsstrahlung-induced pair production.  Staying below 10~MeV ensures no persistent radioactive activation in the target zone and that the radiation level outside the biological shield is safe for public access. An annealed tungsten-mesh assembly placed directly behind the target acts as a positron moderator.  The system produces $5 \times10^{7}$ slow positrons per second, a performance demonstrating that a low-energy electron linac is a superior choice over positron-emitting radioactive sources for high positron flux.  
\end{abstract}


%%\item Research highlight 1
%%\item Research highlight 2


%%\tableofcontents

\end{frontmatter}

%%\linenumbers

\section{Introduction}
\label{Introduction}

An intense positron source is an indispensable constituent of all experimental setups which are used to study antihydrogen, the simplest anti-atom \cite{ AEGIS2015, ALPHA2017, ASACUSA2010, ATRAP2016}. Beyond their importance in antimatter research, positrons have been used for some time in materials science to study lattice defects and electronic structure in metals, semiconductors and other solids~\cite{Tuomisto2013,Krause-RehbergBook,ColemanBook,Schultz1988,Fermi1995} and as a sensitive probe in few-body atomic and molecular physics~\cite{Charlton2000,Surko2005,Gribakin2010}. Positronium has also found application in a variety of fundamental and applied investigations~\cite{JeanBook,Rich1981,Cassidy2018}. Of particular relevance here is its use as a porosity diagnostic in polymers and porous oxides~\cite{JeanBook}, via laser excitation in the elucidation of bound state leptonic physics and for applications in single-shot lifetime spectroscopy and the creation of Rydberg states~\cite{Cassidy2018}. The advent of positron trapping and accumulation~\cite{Murphy1992,Jorgensen2005} has facilitated the development of non-neutral positron plasma technology, which has provided new possibilities to manipulate and better control beam properties~\cite{DanielsonReview}.

In the majority of these studies, and indeed for most modern experiments involving positrons, the ability to produce near-monoenergetic, low-energy beams in vacuum is the enabling technology. In this paper we present a slow positron beam based on a compact linac.  The device provides positrons for the GBAR experiment at CERN, but similar systems could serve as the basis of versatile positron spectrometers for most of the areas of contemporary interest in the field, as outlined above.  

In the following sections we introduce the GBAR project, then describe the experimental setup of the positron source: the linear electron accelerator, the electron target with the positron moderator and the positron beam line.  In Section \ref{RadiationProtection}, \ref{PositronDiagnostics} and \ref{ElectronBackground} we discuss problems and solutions regarding radiation protection, beam diagnostics and electron background.  Finally, we present the currently attained intensity of the positron source and characteristics of the beam in Section \ref{Results}.

\section{Positron source for the GBAR experiment}
\label{GBAR}

The GBAR collaboration aims at a precise measurement of the gravitational acceleration of antihydrogen in the gravitational field of the Earth \cite{GbarProposal, Mansoulie2019}. 
The GBAR  scheme is based on the creation of a positive antihydrogen ion (consisting of an antiproton and two positrons), which can be then sympathetically cooled to low temperature and neutralised by laser photodetachment of one of the positrons. The resulting anti-atom is sufficiently cold for the direct observation of the gravitational free fall.
The anti-ions are created in two consecutive reactions using a dense positronium (Ps) cloud that serves as a target for antiprotons: $\bar p + Ps \rightarrow \bar H + e^{-}$ followed by $\bar H + Ps \rightarrow \bar H^+ + e^-$. In the first step, an antiproton interacts with a positronium to create an antihydrogen atom. Subsequently this newly formed antiparticle reacts with a second positronium and produces a positive antihydrogen ion. The anti-ion formation cross section is thus proportional to the square of the positronium density, itself proportional to the positron flux.

The experiment will receive antiprotons at 100~keV kinetic energy from the new ELENA ring of the AD facility at CERN \cite{ELENA2018}, which are then further slowed down to a few keV energy by an electrostatic decelerator. The expected intensity is $4\times10^6$ antiprotons per pulse.
In order to create one anti-ion, this antiproton pulse must interact with a positronium target of order $10^{10}$ cm$^{-3}$ density.  This high positronium density is obtained by implanting a pulse of the order of $10^{10}$ positrons onto a converter consisting of nanoporous silica. The positronium is formed in an 1 $\mu$m thick porous silica layer. The short lived (125 ps lifetime) spin singlets decay quickly while the long lived (142 ns lifetime) spin triplets are released into the vacuum of a cavity and form the positronium target cloud. As each antiproton must react successively with two positronium atoms to form an anti-ion, the density of the target cloud, hence the positron pulse intensity, is a crucial factor for the success of the experiment.  Positrons are first trapped in a buffer gas accumulator ~\cite{Murphy1992,Amelia2017} and then collected in a high-field (5~T) Penning-Malmberg trap~\cite{Grandemange2013} during the 110 s time lapse between two  antiproton pulses, before being ejected in an intense pulse onto the positron-positronium conversion target.  In order to be trapped the positron energy must be in the eV--keV range (the so called slow positrons).

Low-energy positron generators most often use commercially available $^{22}$Na radioactive sources. Their activity is however limited in practice to approximately 50 mCi (1.7~GBq), which in combination with a solid neon moderator \cite{Cassidy2006, Clarke2006, Cooper2015} can lead to a maximum of $10^{7}$ low-energy positrons per second. Furthermore, the half-life of $^{22}$Na is 2.6 years, thus the low-energy beam intensity reduces over time and the source requires periodic and complicated replacement. Devices using nuclear reactors~\cite{Hugenschmidt2008,VanVeen1999,Liu2013,Sato2015}  or large accelerator facilities~\cite{Wagner2018,Kurihara2000} can potentially provide a much higher positron flux. 
However, for the GBAR experiment, where a high positron intensity is crucial, a dedicated facility, with moderate size and cost, is the only feasible solution. 
We have thus chosen to construct a positron generator based on a low-energy linear electron accelerator because it has the potential to deliver a higher positron flux and at the same time it is compact enough to be placed in the available experimental area at CERN.

\section{Electron linear accelerator}
\label{Linac}

There have been a number of slow positron beam systems which used an electron linac as a positron source.  The performance of a few of them  is listed in Table~\ref{PositronSourceTable}. High-energy electrons hitting a dense metallic target abundantly generate positrons by pair production. However, these particles can only be used for beams after reduction of their energy by a positron moderator.
The efficiency of the slow-positron production in the devices listed, defined as the number of slow positrons per electron impinging on the target, is in the 0.06$\times10^{-7}$--15$\times10^{-7}$ range. Potentially, higher efficiencies can be achieved by increasing the electron energy. However, the actual performance depends on the geometry of the target-moderator structure and efficiency is not the sole design criterion in many high-energy, high-power electron accelerators. Higher energy also requires a longer accelerator structure and a thicker biological shield.  
In the case of the GBAR source in the AD hall, the radiation dose rate outside the biological shield must be compatible with public access. Even with electron kinetic energies as low as 9 MeV, the electron bremsstrahlung radiation produces neutrons by interacting with some nuclei present in the surrounding structural materials via the ($\gamma$,n) reaction. This process leads to the creation of short-lived radioisotopes in the vicinity of the electron target, but the total activation level is low and the target can be approached immediately after switching off the linac. Above 10~MeV activation increases rapidly with energy. Dose-rate simulations showed that the size of the radiation shield at 18~MeV electron beam energy would be incompatible with the available volume and access to the target zone would be severely limited.  

\begin{figure}[]
\includegraphics[scale=0.6]{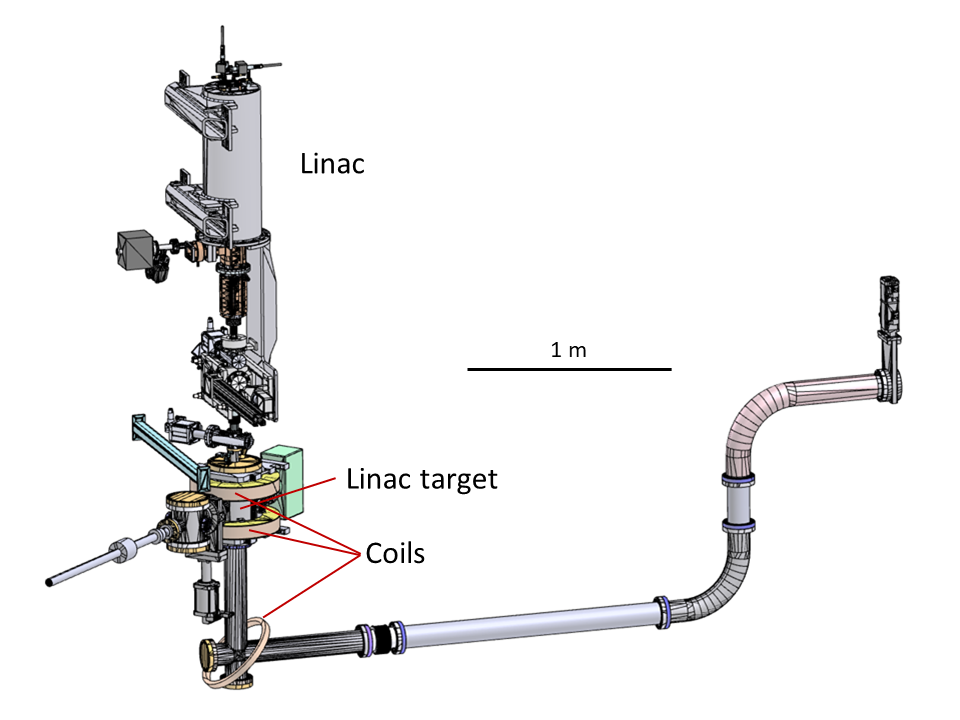}
\caption{Schematic view of the linac (vertical structure on the left) and the positron transfer line. The transfer fields are generated by solenoids wound around the beam pipes and by the two larger coils placed around the linac target. The coil at 45 degrees position is used to fine-tune the magnetic field at the point where the positron beam turns sharply. The beam line crosses the biological shield at the ``S'' shaped section on the right.}
\label{FigurePositronLine}
\end{figure}

The GBAR electron accelerator  (Fig. \ref{FigurePositronLine}) is a water-cooled linac with a thermionic triode cathode, constructed by NCBJ (Poland). The microwave power is supplied by a 7.5~MW klystron (Thales 2157A), regulated by a solid-state  modulator with pulse transformer from ScandiNova Systems. The accelerating section is composed of 18 cavities of 4.5~mm aperture radius with a total  length of 900 mm. It is surrounded by a solenoid which provides a 59~mT longitudinal magnetic field.  The cavity is mounted in a vertical position. The accelerator produces electrons at 9~MeV kinetic energy (with 0.5~MeV FWHM) in 2.85~${\mu}$s long pulses (FWHM). The repetition rate can be varied from 2 to 300~Hz. The peak electron current is 330~mA. The energy distribution has been verified by a magnetic dipole spectrometer. The size and position of the beam spot have been optimised by a removable YAG (yttrium aluminium garnet) screen, placed between the linac and the electron target and observed by a camera. The viewport and the camera of this diagnostic device have to be removed in normal operation, as they cannot withstand the very high radiation dose. Focusing and position of the electron beam can be controlled by a triplet magnetic lens system at the exit of the linac cavity. The beam can be focused to a spot as small as 3~mm in diameter. In the positron-production setting, we slightly defocus the beam spot, to approximately 5~mm diameter, in order to avoid local overheating of the target. The klystron and the accelerating section are both equipped with a closed-cycle water cooling system.

We constructed a test installation in CEA-Saclay~\cite{Liszkay2014} (in the following, we will refer to this system as the ``Saclay source'') which is based on a 4.3~MeV  linac with a magnetron as microwave source. It provides electron pulses with 150~mA peak current, 2.5~$\mu$s pulse length and 200~Hz repetition rate.
The Saclay source produced $2\times10^6$ slow positrons per second, a performance which is comparable with the yield of neon moderated isotope sources.  The results proved that a source based on a low energy linac is a suitable device to supply positrons for the GBAR experiment. Some optimisation of the moderator and the target has been done with this facility.

\section{Linac target}

In electron-linac-based sources, positrons are mostly created by the \hyphenation{brems-strahl-ung} bremsstrahlung radiation emitted by electrons impinging on a dense metallic target with high atomic number. The energy spectrum of positrons generated by electrons at 9 MeV kinetic energy extends to approximately 7 MeV. It must be reduced in order to allow their subsequent transport and trapping. The ``fast" antiparticles are slowed  to an energy of a few electronvolts by a positron moderator \cite{ColemanBook,Schultz1988}.  A high positron intensity requires a large power dissipated in the target, therefore both an efficient target cooling system and a sophisticated moderator configuration are important for high positron yield. 

\begin{figure}[]
\includegraphics[scale=0.6]{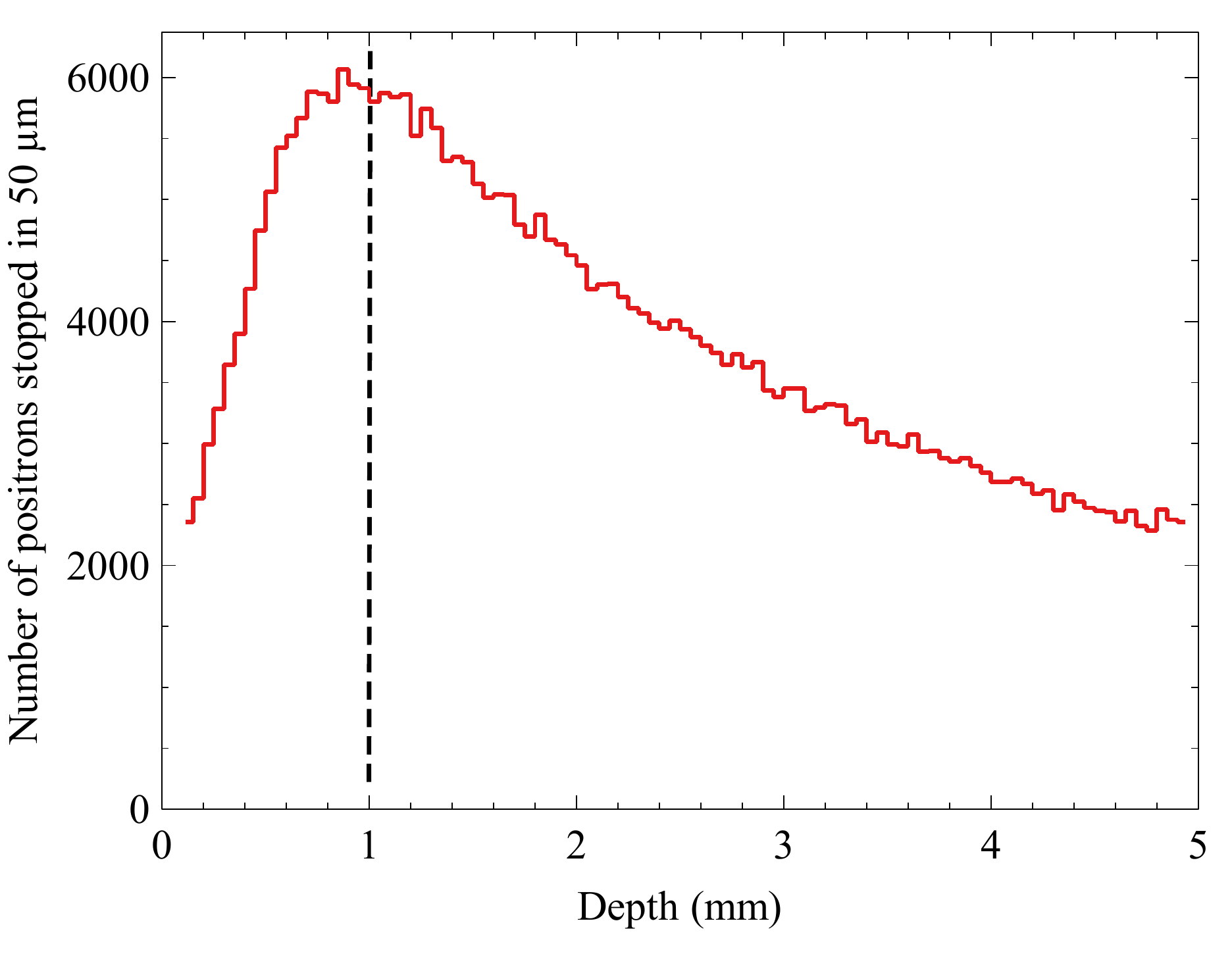}
\caption{Positrons created by 9~MeV electrons and stopped in a tungsten plate (\textsc{Geant4} simulation with $10^7$ electrons). The number of positrons annihilating in 0.05~mm thick layers is plotted as a function of the depth. The dashed line at 1 mm shows the thickness of the actually used electron target.}
\label{FigureStoppingProfile}
\end{figure}

The linac target generates high energy positrons for subsequent moderation. It also produces a high flux of bremsstrahlung radiation which creates positrons in the moderator itself, this latter process being responsible for approximately 40~\% of all slow positron output of the source. The target is made of tungsten, as this metal was found to be the best choice for electro-production of positrons~\cite{Dahm1989} because of its high atomic number and melting point. It has been designed to produce the highest number of slow positrons possible at 9~MeV electron energy while having sufficient cooling power to keep the target at moderate temperature. The thickness of the target has been optimized by simulating the stopping profile of positrons, created by electrons implanted into a thick tungsten plate  (Fig. \ref{FigureStoppingProfile}), using the \textsc{Geant4} simulation toolkit~\cite{GEANT4}. As we use tungsten for both the target and the moderator, the maximum of the stopping profile gives the depth where the highest number of slow positrons are produced in the moderator behind a target of a given thickness. Figure \ref{FigureStoppingProfile} shows that the maximum of slow-positron production efficiency is at approximately 1~mm depth. In the case of the GBAR linac the optimal thickness is used with a simple static construction, without rotating target or scanning of the beam, as the cooling system is able to absorb the power deposited in a 1~mm thick target. Although the Monte Carlo simulation is reliable only down to a few hundred electronvolt positron kinetic energy, the calculated profile is very close to the actual stopping profile because positrons in this energy range do not move more than 100~nm before thermalisation. Consequently, the calculated stopping profile is a good approximation for the depth dependency of positrons which are available for the final phase of moderation. The maximum yield is a broad function of the depth, with marginal changes only at the scale of the thickness of the moderator. The target is perpendicular to the beam axis. To improve heat conduction, it is machined in the form of a 5~mm diameter, 1~mm thick disc, milled out of a thicker tungsten block. It is in turn attached to a water-cooled copper structure. When the linac works at nominal power, the target assembly absorbs 1.5~kW. On the basis of a finite element calculation we estimate the temperature of the target as 1400~K. This high temperature is localized to the electron beam spot, the rest of the target structure is efficiently cooled by the water circuit.

\section{Positron moderator}
In the positron moderation process, high-energy positrons are implanted into a solid where they lose energy until they are close to thermal equilibrium with the crystal lattice \cite{Schultz1988}. Some of the thermalized positrons reach the surface by diffusion. In the case of tungsten and some other materials, the work function of positrons is negative, i.e., the particles gain energy when they leave the metal. Only positrons that reached thermal energy in the vicinity of the surface, in a depth range in the order of 100~nm, have a chance to be emitted from the moderator. Positronium formation and positron-electron annihilation are other possible surface processes which limit the efficiency. Altogether, moderation is a near-surface process and its efficiency increases with the useful emitting area of the moderator structure. However, in complicated moderator structures most slow positrons can only leave the moderator after one or more collisions, with a significant probability of loss at each interaction. Intense sources based on the $^{22}$Na isotope most often use solid neon kept at 7~K as moderator. The high power due to the scattered electron beam and high-energy gamma rays in the target zone makes application of a cryogenic moderator very difficult in the case of accelerator-based systems. In linac-based systems the moderator is usually a structure made of metallic plates or foils, annealed at high temperature. The material chosen is most often tungsten due to its high efficiency and relative stability~\cite{Ebel1987,Howell1987,Hulett1989,Akahane1990,Segers1991,Wada2012,Tanaka1991}.
The number of ``fast'' positrons created in the target increases quickly with electron energy. However, the mean energy of the positrons increases as well, which entails a decrease in moderator efficiency. This effect attenuates the gain that arises from increasing the electron energy.

The simplest type of metallic moderator is a thin foil that must be annealed at high temperature in order to release slow positrons efficiently. High-temperature annealing removes defects and thereby increases the effective diffusion length of low energy positrons in the metal. Furthermore, it cleans surface contamination and reduces the loss of positron emission through positron trapping at the surface or positronium creation.
In order to increase the slow-positron yield, we have chosen a stack of tungsten mesh pieces as a moderator because this is a structure with a large surface per unit mass, readily available and easy to heat using electrical current. Similar structures have been found to have a higher moderator efficiency than a thin tungsten foil~\cite{Nagashima2000, Weng2004, Liszkay2002}. While positrons from a $^{22}$Na source with 180~keV mean kinetic energy are quickly absorbed in a few layers of commercially available mesh, in the case of an electron-beam-based source, the number of moderated positrons emitted per unit area is independent of the depth in the stack of moderator layers, because of the broad stopping profile. Consequently, the advantage of a larger specific surface area is expected to be even more enhanced than in the case of isotope sources where the intensity of fast positrons emitted from the source is quickly attenuated in the moderator stack. Slow positrons emitted from a surface deep in the moderator stack can only escape if they undergo a few collisions with the wires of the mesh.  The loss during the collisions limits the gain attained with an increased number of layers, leading to an optimal thickness. A further factor which influences the efficiency is the temperature of the moderator, which depends on the energy deposited in the moderator, which in turn also depends on the thickness.

\begin{figure}[]
\includegraphics[scale=0.6]{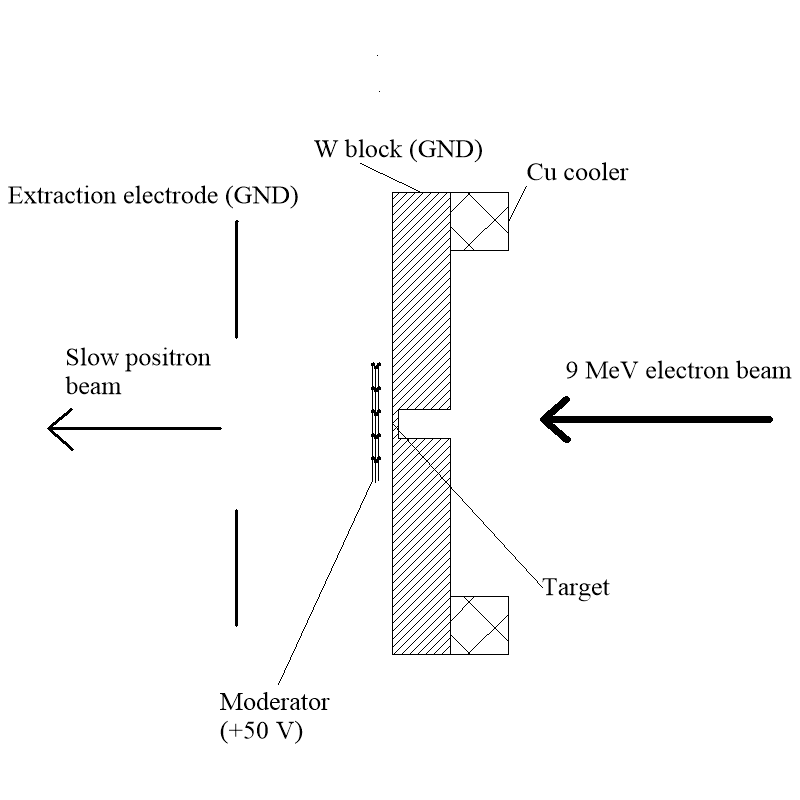}
\caption{Cross section of the electron target. The potential of the moderator is +50~V, the rest of the structure is at ground (GND). The copper block (``Cu cooler'') is water cooled. The magnetic field of 9.7 mT is parallel with the electron and positron beams. }
\label{FigureElectronTarget}
\end{figure}

The stack of thin tungsten mesh pieces is mounted 2~mm behind the electron target (Fig. \ref{FigureElectronTarget}), parallel to the target. A woven wire mesh of 0.0008'' (20.3~$\mu$m) thickness was used with 180 wires/inch density (141~$\mu$m wire distance). The mesh pieces were annealed in a vacuum chamber, with a pressure lower than $10^{-7}$ mbar, using an electrical current giving a power density of about 100 W/cm$^2$. They were mounted on a stainless steel moderator holder in air and moved into the target chamber in typically less than 15 minutes.
The moderator is biased at +50 V. The extraction electrode is a simple grounded plate with a 20 mm diameter hole located 27~mm downstream of the target. This simple design ensures that the electrode is not excessively heated by the electron beam. Simulations have shown that the extraction field does not have a significant negative effect on the beam quality. The slow positron yield was measured by a NaI scintillator coupled to a photomultiplier at the exit point of the biological shield. 

\begin{figure}[]
\includegraphics[scale=0.6]{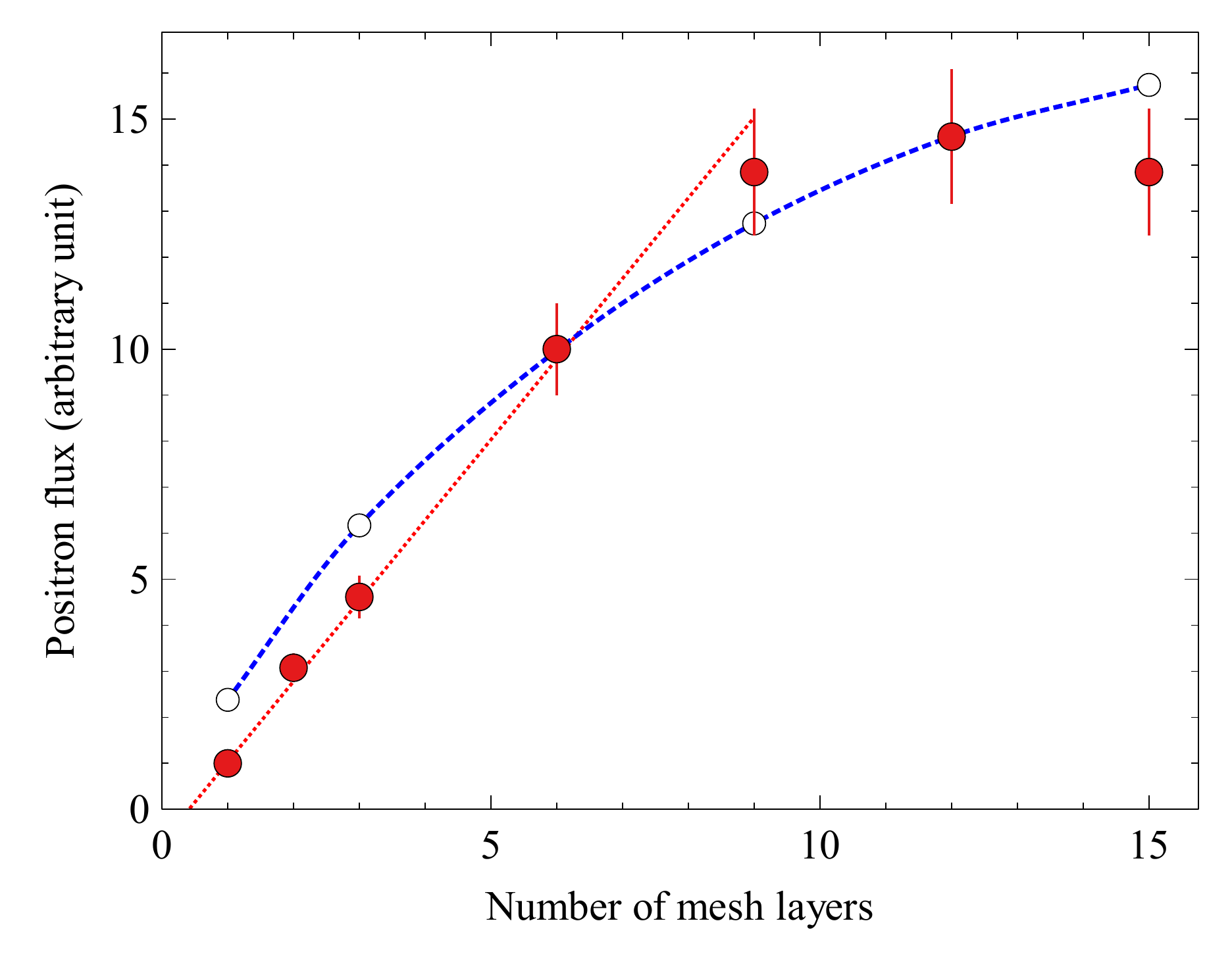}
\caption{Slow positron flux as a function of the number of mesh layers (solid circles). Each layer is a 18x18~mm piece of tungsten mesh with 180 wires/inch density, annealed at higher than 2700~K temperature. The measurement was performed using the 4.3~MeV Saclay linac with a 1~mm thick tungsten target. A linear fit of the data points up to 9 layer thickness is shown with a dotted line. Results of a simple simulation (see text) are displayed with open circles. The dashed line is guide for the eye.}
\label{FigureMultipleMesh}
\end{figure}

We performed measurements using the Saclay source to optimize the positron moderator. Its ideal thickness was obtained by measuring the slow-positron yield as a function of the number of mesh layers in the moderator stack (Fig. \ref{FigureMultipleMesh}). The experimental uncertainty is dominated by the difference in the properties of the electron beam between two measurements, as changing the moderator thickness is a time consuming procedure and each measurement was performed after a new start of the accelerator. We estimate this error to approximately 10~\% of the signal. We found that the positron yield increases nearly linearly with the number of mesh layers up to about 9 layers, then the signal levels off. We concluded that the optimal thickness of the moderator is approximately 12 layers. In the figure, results of a simple Monte Carlo simulation are also shown, using a probability of slow positron reflection from the surface of the wire of 0.56 \cite{Jorgensen1998}. The simulation was normalized to give the same efficiency value as the measurement at a thickness of 12 layers. As it is also visible in the simulation, we expect a small increase in positron yield between 12 and 15 layers. The apparent decrease in the measured values is compatible with the experimental uncertainty. Nevertheless, no significant improvement can be expected by increasing the number of mesh layers beyond 12 layers. As the change in the stopping profile of positrons within the thickness of the moderator is small, the density of positrons created per unit surface area is nearly independent of the depth within the moderator. Consequently, if the moderator efficiency per unit surface is unchanged we expect that the optimal thickness is the same at 4.3 MeV and 9 MeV electron energy, for the Saclay source and at CERN, respectively. This approximation does not take into account the loss of efficiency due to the increase in temperature with increasing moderator thickness, an effect which is significant in the case of the CERN beam (see below and \ref{PositronFlux}).

We compared the efficiency of the optimized moderator stack with that of a simple flat moderator, placed in the same position. We used pieces of 25~$\mu$m thick and 1~mm wide tungsten ribbon to construct a flat moderator. This geometry allows heating the metal by electrical current in the same chamber as done for the tungsten mesh. The difference between the geometry of the ribbon and that of the tungsten-mesh moderator was taken into account using a Monte Carlo simulation of the target-moderator structure. The simulation provides the density distribution of moderated positrons in the plane of the moderator. The efficiency of the flat moderator was experimentally found to be only $17 \pm 5$~\% of the efficiency of the optimized mesh configuration, which confirms the expectation that mesh moderators have significantly higher efficiency than thin foils, particularly, as here, for positrons incident with kinetic energies in the MeV range.

\begin{figure}[]
\includegraphics[scale=0.6]{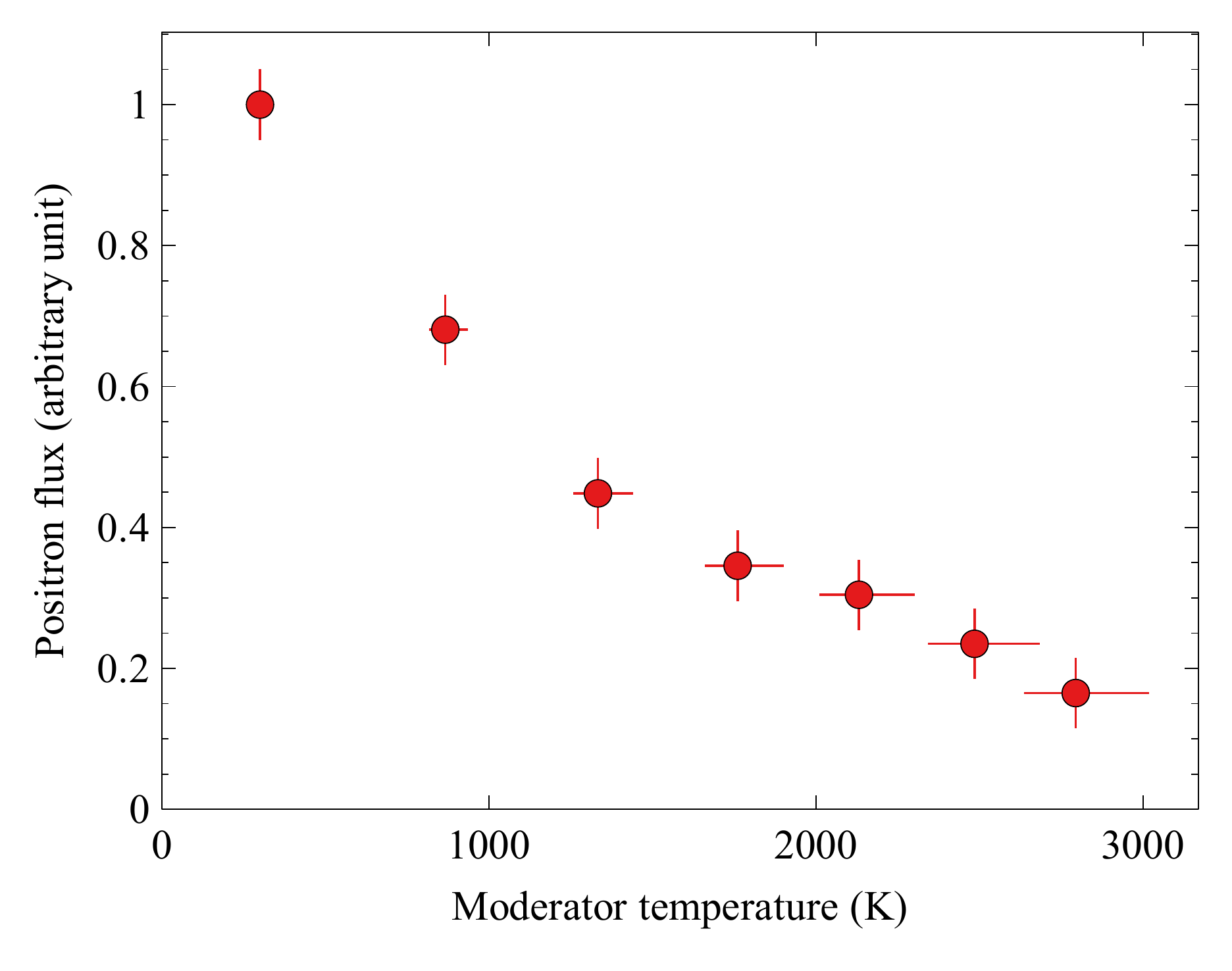}
\caption{Slow positron flux as a function of the moderator temperature. The measurement was performed using the 4.3 MeV Saclay linac with 1 mm thick tungsten target. The temperature was estimated on the basis of the heating power and radiative heat exchange.}
\label{FigureInSituHeating}
\end{figure}

We also studied the effect of the temperature on the moderator efficiency at the Saclay source using \emph{in-situ} heating of a single moderator mesh by electrical current (Fig. \ref{FigureInSituHeating}). The moderator was placed just behind the electron target, in a similar position as the standard moderator stack used in the setup. We estimated the temperature of the moderator on the basis of the heating power and radiative heat transfer, using an emissivity of 0.3 for the tungsten wire. As the exact value of the emissivity of the used wire is not known, the experimental error of the temperature is rather large. In the figure, the error bar represents the range of temperature determined expecting an emissivity between 0.2 and 0.4. The uncertainty in the positron signal is limited to small fluctuations in the electron beam intensity during the measurement. We found that the moderator efficiency decreases by as much as 30~\% between room temperature and 800~K, then continues to decrease more slowly dropping to about 15~\% at 2800~K. The result is in qualitative agreement with the measurements of Al-Qaradawi et al~\cite{Coleman2002}. The loss of efficiency can be attributed to the increasing positronium formation at the moderator surface at elevated temperature. As positronium formation competes with the emission of slow positrons from the surface of the tungsten mesh, this leads to a decrease in moderator efficiency. 

\begin{figure}[]
\includegraphics[scale=0.6]{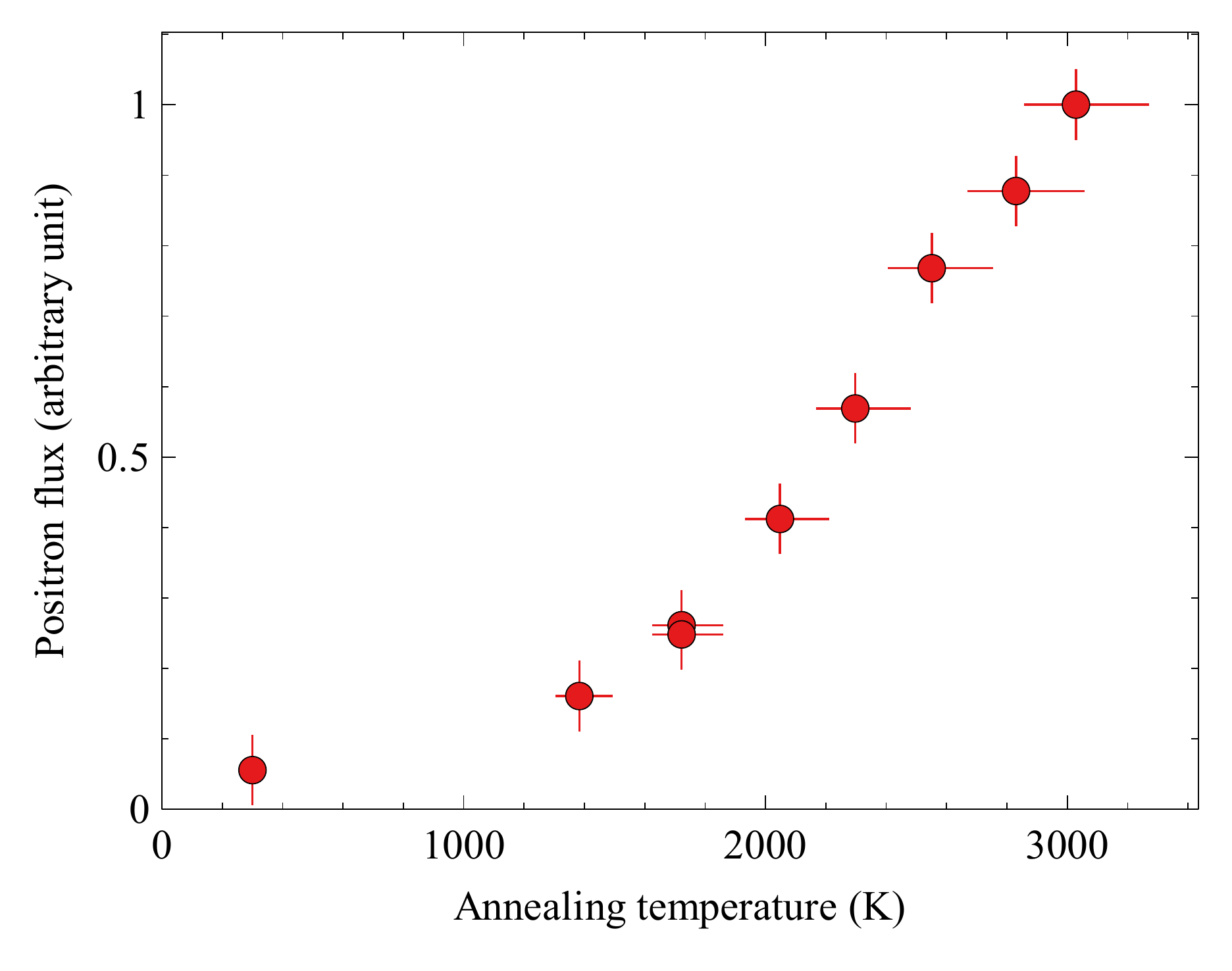}
\caption{Slow positron flux as a function of the moderator annealing temperature. The measurement was performed using the 4.3 MeV Saclay linac with 1 mm thick tungsten target. The temperature was estimated on the basis of the heating power and radiative heat exchange.}
\label{FigureInSituAnnealing}
\end{figure}

We performed \emph{in-situ} annealing of the moderator mesh using the same experimental setup. The intensity of the slow positron beam was measured after heating the mesh for 3 minutes at various heating power levels. Figure \ref{FigureInSituAnnealing} shows the positron flux as a function of the estimated annealing temperature. The experimental uncertainties are the same as in the case of Figure \ref{FigureInSituHeating}. The moderator efficiency is very low in the case of an unannealed mesh. The positron signal starts to increase above 1800~K annealing temperature and increased roughly linearly until 3000 degrees. This result illustrates the importance of annealing of the moderator at the highest temperature which is technically possible.

\section{Positron transport}
\label{PositronTransport}

The target is surrounded by two coils in the Helmholtz configuration arranged co-axially with the same axis as that of the accelerating section. They produce a field that can be varied up to about 20 mT at the target location. Positrons are adiabatically guided by an 8~mT magnetic field which is generated by solenoids wound around the 100~mm diameter vacuum pipe (Fig. \ref{FigurePositronLine}). At vacuum valves, bellows and other vacuum elements, larger coils are used to provide a smooth field. At each elbow, two pairs of racetrack coils introduce a variable dipole field that can be used for steering the positron beam. The total length of the beam transfer line between the electron target and the entry point of the positron trap is approximately 7.5~m. The ``S'' shaped part of the beam trajectory before the vacuum valve is in the zone where the beam line crosses the biological shield and a reinforcement of the radiation shield is necessary to avoid leakage of gamma radiation from the linac bunker.

\section{Radiation protection}
\label{RadiationProtection}

The intense electron pulses produce a very high radiation dose rate (up to 30 kGy/h) in the vicinity of the target chamber. Thus, the linac and the target chamber are placed in a bunker of approximately $10\times11$~m footprint with 1.2~m thick walls, constructed from 67~\% concrete and 33~\% iron blocks. A stainless steel shielding box, with 40 mm thick walls,  has been installed around the target, to protect the equipment in the linac bunker. This reduces the radiation dose by about a factor of 3 inside the bunker, and consequently outside. The radiation dose rate outside the bunker is sufficiently low for unlimited access. This means that it might be possible to contemplate the use of this relatively compact type of instrumentation in small laboratories. We observed a slight short-term activation of the mechanical structure around the linac target but it never exceeded a few $\mu$Sv/h equivalent dose rate at 400~mm distance from the target.  Use of lead as shielding material is prohibited at this energy, as it would significantly increase the level of activation. 

\section{Positron diagnostics}
\label{PositronDiagnostics}

In order to measure the positron flux at the end of the transport line~\cite{Latacz2019}, a 0.5 mm thick stainless steel plate is used as a beam diagnostic target. 
The target is mounted on a linear drive and can also be used as a scraper to estimate the beam size. A NaI(Tl) scintillator (50.8~mm diameter, 50.8~mm length) attached to a photomultiplier tube is used to detect 511~keV photons from positrons annihilating on the diagnostic target. The detector is placed at 800~mm distance from the target, so that the systematic error caused by uncertainties of the geometry can be neglected and calibration with single annihilation gamma photons can be used. To calibrate the detector efficiency, we first measured the mean electrical charge corresponding to the 511~keV photopeak at the anode of the photomultiplier ($Q_{511keV}$). In a second step we determined the mean energy $E_{a}$ that is deposited in the scintillator crystal after annihilation of one positron in the target. This was done by a Monte Carlo simulation of the detection using the \textsc{Geant4} package~\cite{GEANT4}. The simulation takes into account scattering in the environment of the positron target (vacuum pipe, target holder, vacuum flanges), the solid angle of the detector and Compton scattering in the scintillator. The charge corresponding to the photopeak is corrected by the factor determined in the simulation to obtain the mean charge $Q_{m}$
\begin{equation}
Q_{m}=Q_{511keV}\frac{E_{a}}{511\ keV}
\end{equation}
from the annihilation of one positron in the target. At the detector distance used, each positron pulse produces on average only a few  tens of 511 keV annihilation photons that reach the detector. This allows the determination of both the single-photon signal and measurement of the diagnostic signal with a moderate dynamic range.
 
\begin{figure}[]
\includegraphics[scale=0.6]{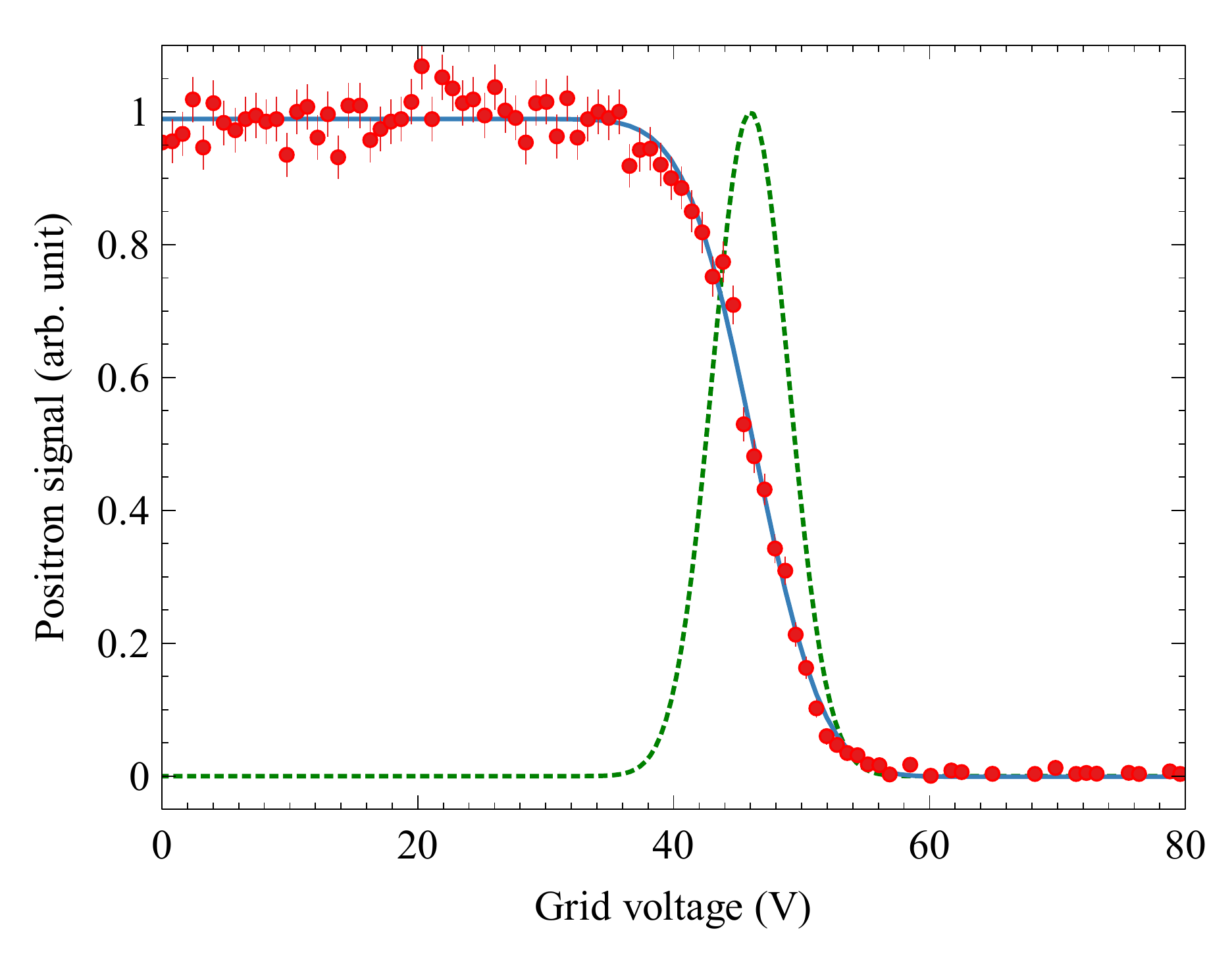}
\caption{Positron annihilation signal as a function of the voltage of the grid of the retarding field analyzer at 50~V moderator voltage. At low grid potential all positrons are annihilated on the target while above approximately 60~V all particles are repelled by the grid and no positron signal is detected. The continuous line is a fit with a complementary error function, giving $\sigma_{\parallel t}=$4.2~eV. The dotted line shows the corresponding Gaussian energy distribution. The measurement was performed in a 60~mT longitudinal magnetic field.}
\label{FigureRetardingFieldSignal}
\end{figure}

In order to measure the momentum distribution of the positrons parallel to the beam axis, we used the first electrodes of the buffer-gas trap as a retarding field analyser. The relevant part of the trap consists of a series of tubular electrodes with an inner diameter of 16 mm and a total length of 149 mm.  In this case the positron flux was measured by detecting the annihilation gamma signal generated by positrons impinging on a target behind the last electrode by a plastic scintillator coupled to a photomultiplier. The trap is directly behind the positron diagnostic target and the magnetic field is 60~mT at the place of the measurement. The tubular electrodes are sufficiently long to ensure that the electric potential at the center is equal to that of the electron tube. The energy distribution can be deduced from the electrode voltage - positron annihilation signal curve. A fit of the measured values with a complementary error function shows a good agreement with $\sigma_{\parallel t}=$4.2~eV fitting parameter, corresponding to a Gaussian energy distribution with $\sigma_{\parallel t}$ standard deviation (Fig. \ref{FigureRetardingFieldSignal}).

\section{Electron background}
\label{ElectronBackground}

Electrons are generated in the linac target chamber with a wide range of kinetic energies. Low-energy electrons are adiabatically guided together with the positron beam. Below $\sim$100 eV the electron background is several orders of magnitude stronger than the positron flux and we observe a significant number of electrons even above 2~keV kinetic energy. This background is not noticed in most applications (positron spectroscopy) but gas ionisation by electrons is potentially deleterious in buffer-gas traps. Only a potential barrier of about -5~kV can fully eliminate the electron background. A high-transparency (90~\%) metallic grid at negative potential is used to block most electrons. Acceleration and subsequent deceleration of positrons by the grid leads to a deterioration of the beam quality, therefore the potential on the grid must be limited to the lowest possible level. We found that the best trapping efficiency in the buffer gas trap is attained for a grid potential of -500~V.

\section{Results}
\label{Results}

\subsection{Positron flux}
\label{PositronFlux}
\begin{figure}[]
\includegraphics[scale=0.6]{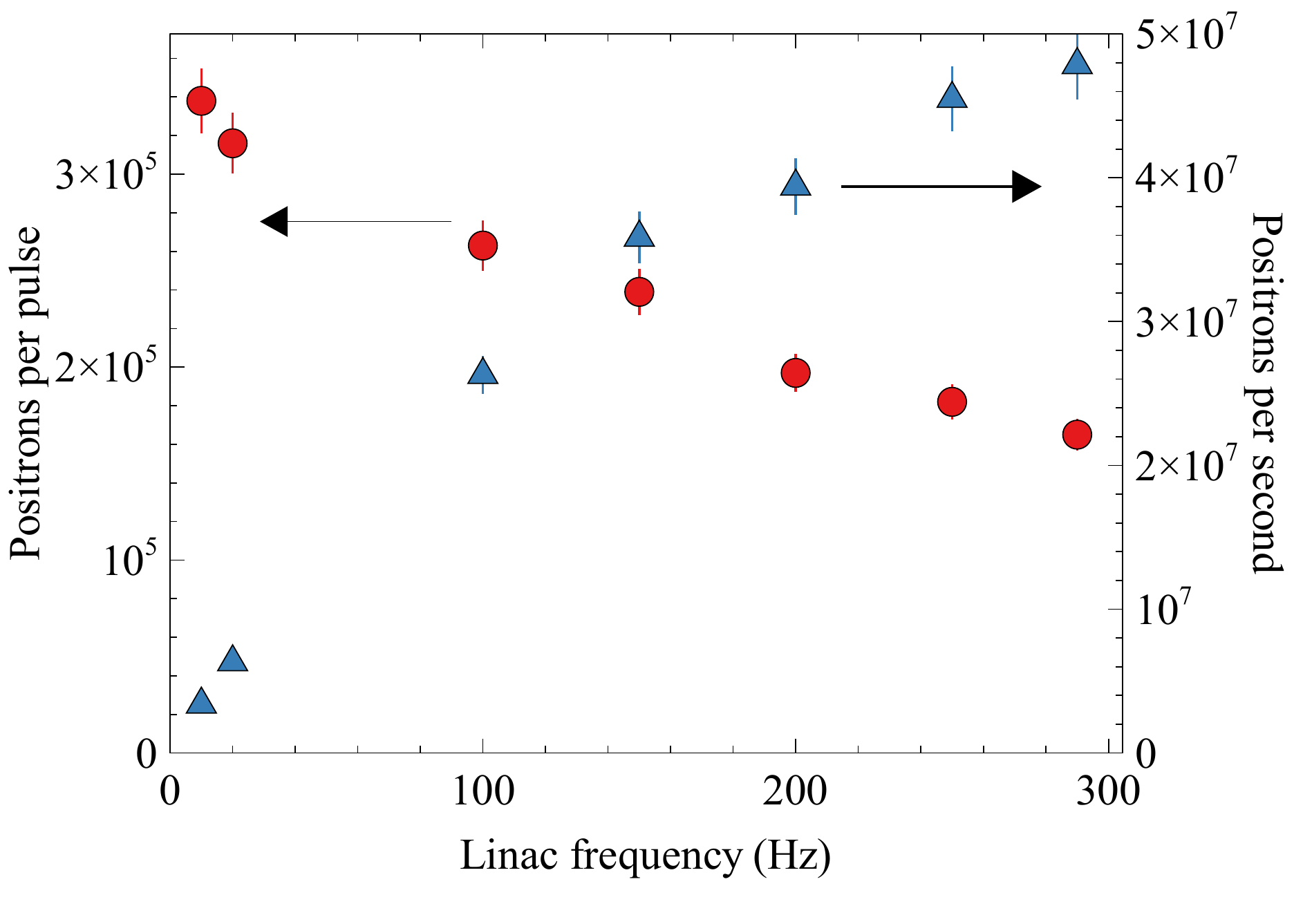}
\caption{Slow positron yield of the GBAR positron source as a function of the linac frequency. Both the number of positrons per pulse (circles) and the number of positrons per second (positron flux) (triangles) are shown. The yield was measured after more than 30 minutes operation at a given frequency.}
\label{FigureFrequency}
\end{figure}

A steady-state positron flux of $5.0 \pm 0.6 \times 10^{7}$ per second was detected at the diagnostic target at 300~Hz repetition frequency \cite{Latacz2019}. The number of positrons per linac pulse measured as a function of the repetition rate is shown in Figure \ref{FigureFrequency}. It decreases almost linearly by about 50~\% between 10~Hz and 300~Hz, most probably due to the increase of the moderator temperature with increasing linac frequency. The uncertainty of the positron intensity measurement is limited to the effect of small fluctuations in the beam intensity, estimated to 5~\%.
The moderator is heated directly by the electron beam (estimated as 150~W at full power by simulation) and indirectly by thermal radiation of the linac target. With no efficient cooling by heat conduction, its temperature is determined by radiative equilibrium. 

The positron source has been running at full power for an extended period of time (more than 1000 hours). After installing a fresh moderator, the slow-positron yield stabilized after a short transition period (typically  a few hours). On a longer time scale, there is a slow deterioration with accumulated electron dose. The long term deterioration of the positron yield can be attributed to both surface contamination and accumulation of lattice defects ~\cite{Suzuki1998}.

\subsection{Energy distribution and beam shape}
\label{EnergyDistribution}

Tungsten mesh moderators are characterised by a rather broad angular distribution of the emitted positrons due to the microstructure of the moderator stack. This leads to a longitudinal momentum distribution $p_{\parallel}$ that can be translated into an energy $E_{\parallel} = \frac{p_{\parallel}^{2}}{2 m} $ with a total width of approximately 3~eV, the work function of tungsten. The broadening may be slightly increased by the electric field which penetrates into the mesh stack and may extract some positrons which are emitted backwards. The $E_{\parallel}$ distribution measured by the energy analyzer depends on the magnetic field (B) at the location of the moderator and of the energy analyser. $E_{\parallel}$ measured at 60~mT magnetic field (Fig. \ref{FigureRetardingFieldSignal}) can be fitted by a Gaussian with $\sigma_{\parallel t}=$4.2~eV. Assuming a fully adiabatic beam transport this width translates to $\sigma_{\parallel m}=$0.7~eV  (1.6 eV FWHM) in the 9.7~mT magnetic field at the position of the moderator, using the  fact that $E_{\perp}/B_{\parallel}$ is an adiabatic invariant and the total kinetic energy $E=E_{\parallel}+E_{\perp}$ is constant. The latter assumption is only approximately fulfilled due to the non-zero width of the energy distribution of positrons emitted from the tungsten surface. The width of the energy distribution at the moderator is comparable with the 2.1 eV FWHM value found using a $^{22}$Na positron source \cite{Sueoka2003}.

\begin{figure}[]
\includegraphics[scale=0.6]{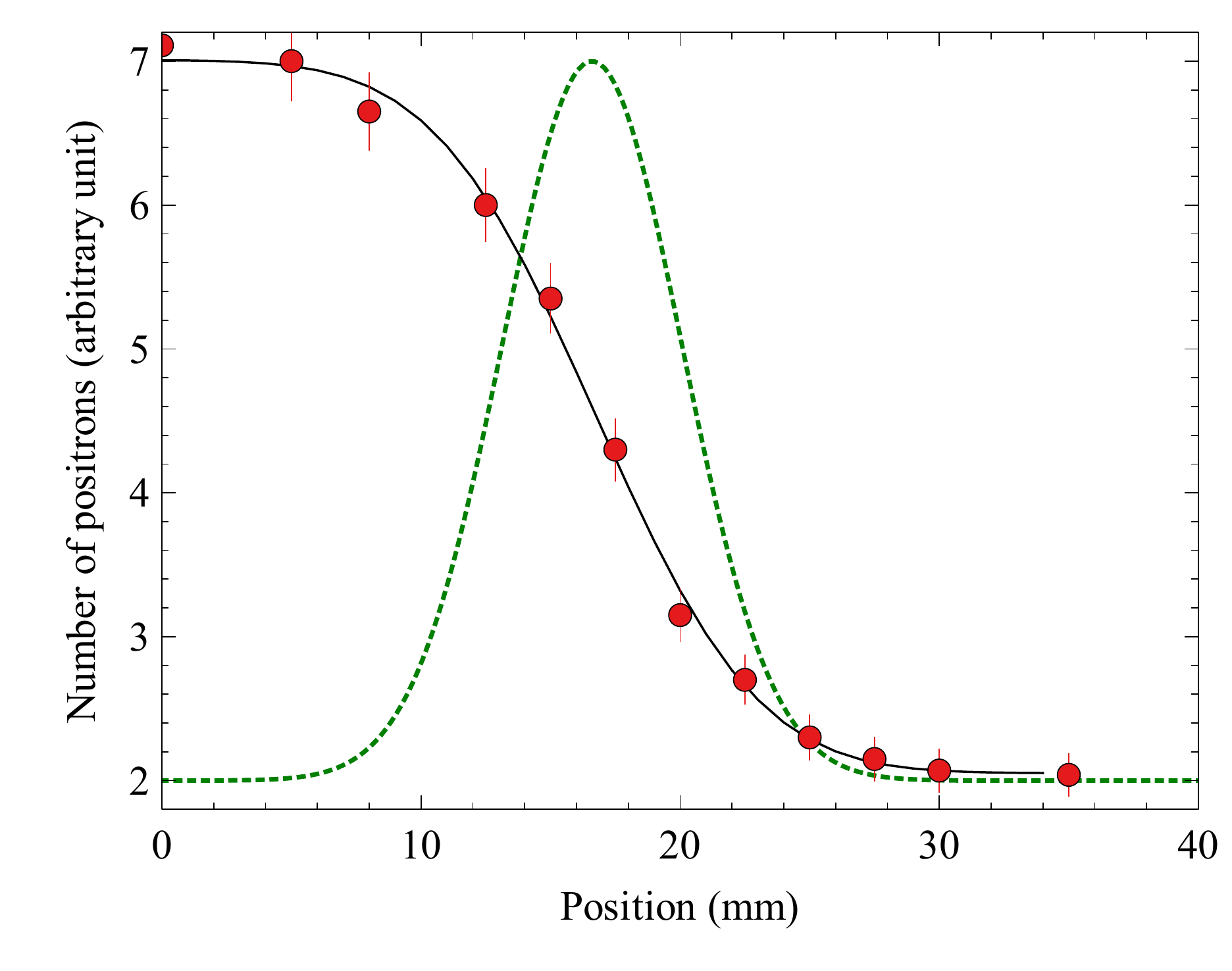}
\caption{Positron annihilation signal as a function of the position of the scraper target. The continuous curve represents a complementary error function fit with $\sigma=4.9$ mm. The dotted line is the corresponding beam profile.}
\label{FigureBeamSize}
\end{figure}

The beam diameter at the positron target can be estimated from measurements with the beam scraper (Fig. \ref{FigureBeamSize}). We found that approximately 80~\% of the positron intensity falls into a 13~mm broad vertical zone. In the case of adiabatic transport the beam diameter depends on the size of the positron emitting spot on the moderator and strength of the magnetic field at both the moderator and the beam scraper. The observed size is in agreement with the expected size of the positron emitting surface of the moderator stack. 

\section{Conclusions and outlook}
\label{Conclusions}

We built and successfully commissioned a positron generator which is based on a compact, low-energy linear electron accelerator. The system provides $5.0 \pm0.6 \times 10^{7}$ e$^+$/s positron flux, which is fed into a buffer-gas trap. The positron flux reached is comparable to or higher than most linac-based positron beams which use significantly higher electron energy. We have demonstrated that a low-energy linac, with no persistent activation of the environment, is a good alternative to radioactive sources when a high positron flux is needed, and as such may find wide uptake. Compared to other linac-based sources (Table~\ref{PositronSourceTable}) the GBAR source provides excellent flux for its input power and energy. The  source will be the first of its kind to be used to fill a high-field Penning-Malmberg trap and it can also serve as a test bench for the application of positron traps at accelerator-based positron beams.

\section{Acknowledgements}
\label{Acknowledgements}

We acknowledge support of the Agence Nationale de la Recherche (project ANTION ANR-14-CE33-0008), CNES (convention number 5100017115), the Swiss National Science Foundation (Grant number 173597), ETH Zurich (Grant number ETH-46 17-1), the Ministry of Education of the Republic of Korea and the National Research Foundation
of Korea (NRF-2016R1A6A3A11932936, NRF-2016R1A2B3008343). Laboratoire Kastler Brossel (LKB) is a \textit{Unit\'e Mixte de Recherche de Sorbonne Universit\'e, de ENS-PSL Research University, du Coll\`ege de France et du CNRS n$^{\circ}$~8552}. We thank to CERN for the support to construct the linac and its biological shield and for fellowships and Scientific Associateship provided to D. P. van der Werf and P. Pérez. The support of the Enhanced Eurotalents Fellowship programme to D. P. van der Werf is acknowledged.  We gratefully acknowledge the help of François Butin and the CERN management and teams that were involved in this project.

\footnote{\copyright 2020. This manuscript version is made available under the CC-BY-NC-ND 4.0 license http://creativecommons.org/licenses/by-nc-nd/4.0/}

\newpage



\bibliographystyle{elsarticle-num} 
\bibliography{GBAR_LINAC_Positron_resubmission_v1}


%\begin{thebibliography}{00}

%%\bibitem[APL_40_751]{test}

%%\bibitem{}

%\end{thebibliography}
\newpage

\begin{table}[]
\begin{center}
\begin{tabular}{ l | r  r  r r r  }		
Linac &  e$^{-}$ energy &   e$^{-}$ beam power & slow e$^{+}$ flux &  efficiency \\
&   MeV & W & $10^{7}$ e$^{+}$/s & $10^{-7}$ e$^{+}$/e$^{-}$\\
\hline
Oak Ridge \cite{Hulett1989}& 180 & 55000 & $10$ & $0.53$ \\
Livermore \cite{Howell1987} & 100 & 11000 & $1000$ & $16$   \\
ETL, Japan \cite{Akahane1990}& 75 & 300 & $1.0$ & $6$ \\
%%Rossendorf \cite{Ebel1987}& 35 & 1.6 mA & $6\times10^{8}$ e$^{+}$/s   \\
KEK \cite{Wada2012} & 55 & 600 & $5$ & $7.3$\\ 
Ghent \cite{Segers1991}  & 45 & 3800 & $2$ & $0.4$\\
Giessen \cite{Ebel1987}& 35 & 3500 & $1.5$ & $0.2$  \\
Mitsubishi, Japan \cite{Tanaka1991} & 18 & 16 &$ 0.077$ & $1.35$\\ 
%%Beijing \\
GBAR, CERN & 9 & 2500 & $5$ & $0.28$  \\
Saclay, CEA \cite{Liszkay2014} & 4.3 & 300 & $0.2$ & $0.05$  \\

\end{tabular}
  \caption{Performance of linac-based positron sources.}  
\label{PositronSourceTable}
\end{center}
\end{table}

\section{Figures}

\end{document}